\documentclass[journal=jpc]{achemso}

\usepackage{titlesec}
\usepackage{indentfirst}
\usepackage{hyperref}
\usepackage{siunitx}
\usepackage{amsmath}
\usepackage[normalem]{ulem}
\usepackage{soul}

\title{Voltage dependent first-principles barriers to Li transport within Li ion battery Solid Electrolyte Interphases}

\author{Quinn T. Campbell}
\affiliation{Center for Computational Research, Sandia National Laboratories, Albuquerque, NM, USA}
\email{qcampbe@sandia.gov}
\date{\today}

\begin{document}

\begin{tocentry}
	\label{TOC Graphic}
	\includegraphics[width=3.3in]{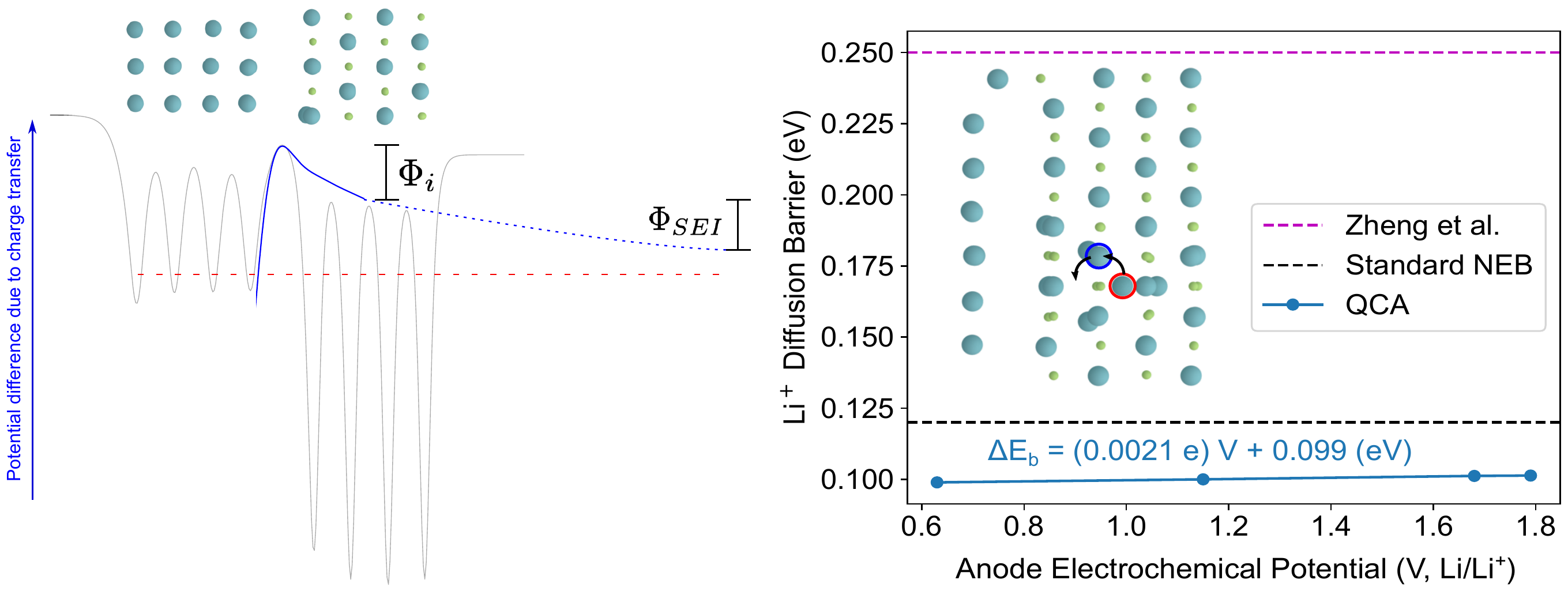}
\end{tocentry}


\abstract{
Charging a Li ion battery requires Li ion transport between the cathode and the anode. 
This Li ion transport is dependent upon (among other factors) the electrostatic environment the ion encounters within the Solid Electrolyte Interphase (SEI), which separates the anode from the surrounding electrolyte.
Previous first principles work has illuminated the reaction barriers through likely atomistic SEI environments, but has had difficulty accurately reflecting the larger electrostatic potential landscape that an ion encounters moving through the SEI. 
In this work, we apply the recently developed Quantum Continuum Approximation (QCA) technique to provide an equilibrium electronic potentiostat for first-principles interface calculations.
Using QCA, we calculate the potential barrier for Li ion transport through LiF, Li$_2$O, and Li$_2$CO$_3$ SEIs along with LiF-LiF, and LiF-Li$_2$O grain boundaries, all paired with Li metal anodes. 
We demonstrate that the SEI potential barrier is dependent on the anode electrochemical potentials in each system.
Finally, we use these techniques to estimate the change in the diffusion barrier for a Li ion moving in a LiF SEI as a function of anode potential.
We find that properly accounting for interface and electronic voltage effects significantly lowers reaction barriers compared to previous literature results. 
}



\maketitle

\section{Introduction}\label{sec1}
Li ion  battery performance is crucial to meeting the world's growing needs to decarbonize transportation \cite{thackeray2012electrical}.
In addition to low weight and volume requirements, faster charging is an important factor in rendering electric vehicles competitive across broader use cases \cite{meintz2017enabling}. 
The rate of charging is determined by a number of factors in the overall design of the battery cell, but is fundamentally determined by the speed at which Li ions can be transported from the cathode to the anode \cite{cai2020review}.
The transport of an individual Li ion is dependent on both the atomistic and electrostatic environment it sees at any given location. 
These can vary significantly across the cell, with one notable bottleneck being Li ion transport through the (typically insulating) Solid Electrolyte Interphase (SEI) layer that develops between the anode and the electrolyte \cite{liu2019challenges}. 
It has been experimentally difficult to characterize the SEI at sufficient detail to understand the precise atomic environment experienced by Li ions. 

First-principles computational work has been crucial to developing an understanding of Li ion transport through SEIs \cite{urban2016computational,wang2018review}.
Due to computational limits, the majority of work in this field has focused not directly on the interface, but on finding reaction barriers in bulk versions of common SEI's such as LiF, Li$_2$O, and Li$_2$CO$_3$ \cite{shi2013defect}. 
More recent work has begun to focus on the direct interface between the anode and the SEI, finding that this interface can be the site of significant changes to the reaction barriers compared to bulk and is crucial for a full understanding of battery behavior \cite{zhang2020stability,leung2020dft}. 
Additional work has also examined the role of grain boundaries in the SEI, finding that these play a potentially significant role in reducing Li ion transport barriers and thus likely serving as the main route for Li transfer \cite{ramasubramanian2019lithium,leung2017spatial,smeu2021electron}.

\begin{figure}[t]
    \centering
    \includegraphics[width=\columnwidth]{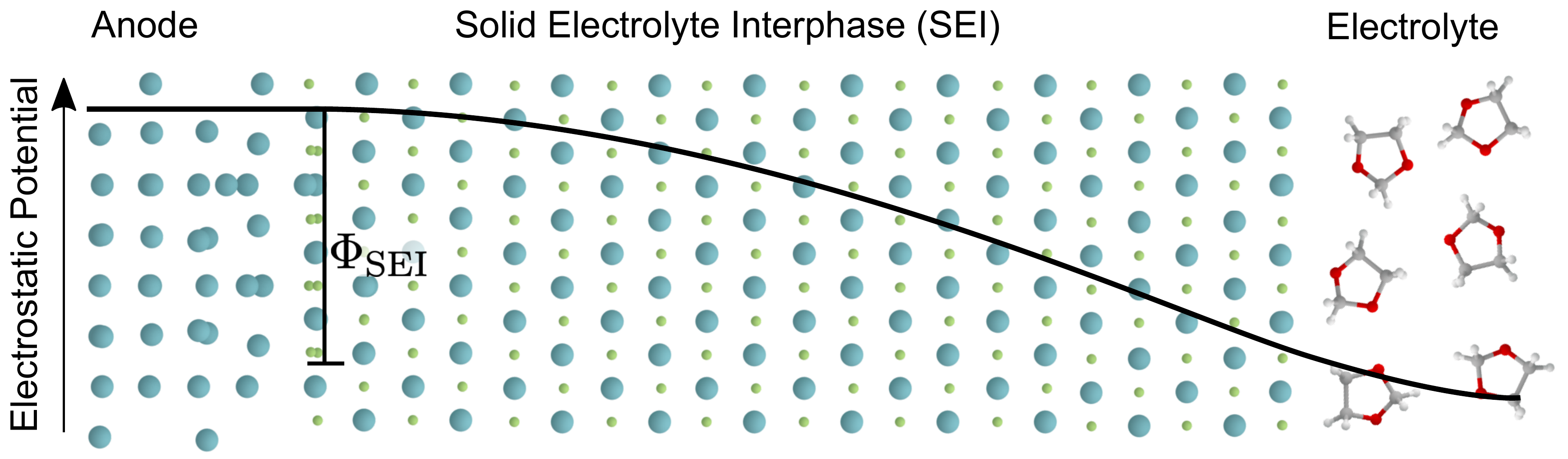}
    \caption{Li transport is highly influenced by both the atomic and electrostatic potential environment the ion encounters. 
    An example of a typical electrostatic potential profile is shown with the dark black line. 
    A Li interstitial's transport barrier near the anode (position 1), within the bulk SEI (position 2), or near the electrolyte (position 3) will be different based on the electric field at these locations. 
    Furthermore, the electric fields across the interface are determined by the equilibrium electronic voltage of the system.}
    \label{fig:schematic}
\end{figure}

While this work has significantly addressed the impact of the atomistic environment on Li ion transport, the electrostatic potential environment has been less thoroughly examined.  
As demonstrated in Fig.~\ref{fig:schematic}, the electrostatic potential experienced by a given Li ion is dependent on its location within the extended battery interface. 
This potential difference across the length of the SEI can give rise to a significant potential barrier that a charge carrier must travel through to move from the electrolyte to the anode.
%
The exact height and spatial location of the barrier are significantly modified by interfacial atomic interactions between the anode and SEI, which need quantum mechanical determination for each specific interface and voltage condition \cite{tung2014physics}, making the height and distribution of barriers \textit{inherently voltage dependent.}
These interfacial barriers have indeed been observed to change with voltage in solid state battery experiments \cite{yamamoto2010dynamic,masuda2017internal}, but remain largely unexplored in non-solid state batteries due to the difficulty of \textit{in-situ} characterization. 

Predicting the barriers for Li ion transport across a realistic SEI electrostatic environment thus requires a quantum mechanical treatment of the anode-SEI interface that is sensitive to the macroscopic potential of the system. 
The typical tool for first-principles quantum mechanical calculations, Density Functional Theory (DFT), however, is currently computationally limited to simulation cells of $\approx$100s to 1000s of atoms.
This is insufficient to fully capture the real world length scale of electrostatics. 
By ignoring the larger interface, however, first-principles calculations of interfacial energetics are inherently only accurate under conditions of zero applied electric field.
Techniques for understanding interfacial barriers must utilize multiscale capabilities that combine DFT with larger analytic potential distributions of bulk materials. 

In this work, we use the recently developed Quantum Continuum Approximation (QCA) technique \cite{campbell2023voltage} to predict the first-principles potential barriers for Li ion transport as a function of both the atomistic and electrostatic potential environment.
QCA works by coupling explicit DFT calculations of relevant interfaces (e.g. anode-SEI, SEI-electrolyte) to Poisson-Boltzmann distributions of charge in the bulk insulating SEI region. 
This allows us to extract potential barriers at electric fields relevant to battery charging operation and assign equilibrium electronic voltages to our DFT calculations.
We examine the barrier for Li ion transport through LiF, Li$_2$O, and Li$_2$CO$_3$ SEI environments as well as LiF-LiF, and LiF-Li$_2$O grain boundaries with a metal Li anode.
We demonstrate that these interfacial barriers are highly dependent on the voltage at which the battery is being charged as well as the atomic interactions of the interface.
We then examine the impact of these SEI barriers on the reaction barrier magnitude of a Li ion moving a single lattice site via the knock off mechanism in a LiF SEI as a function of voltage.
Our results suggest that properly accounting for voltage and interface effects may significantly lower calculated Li diffusion barriers. 

\section{Methods}\label{sec:methods}

\subsection{Determining Potential Gradients across battery SEIs}
\begin{figure}
    \centering
    \includegraphics[width=\columnwidth]{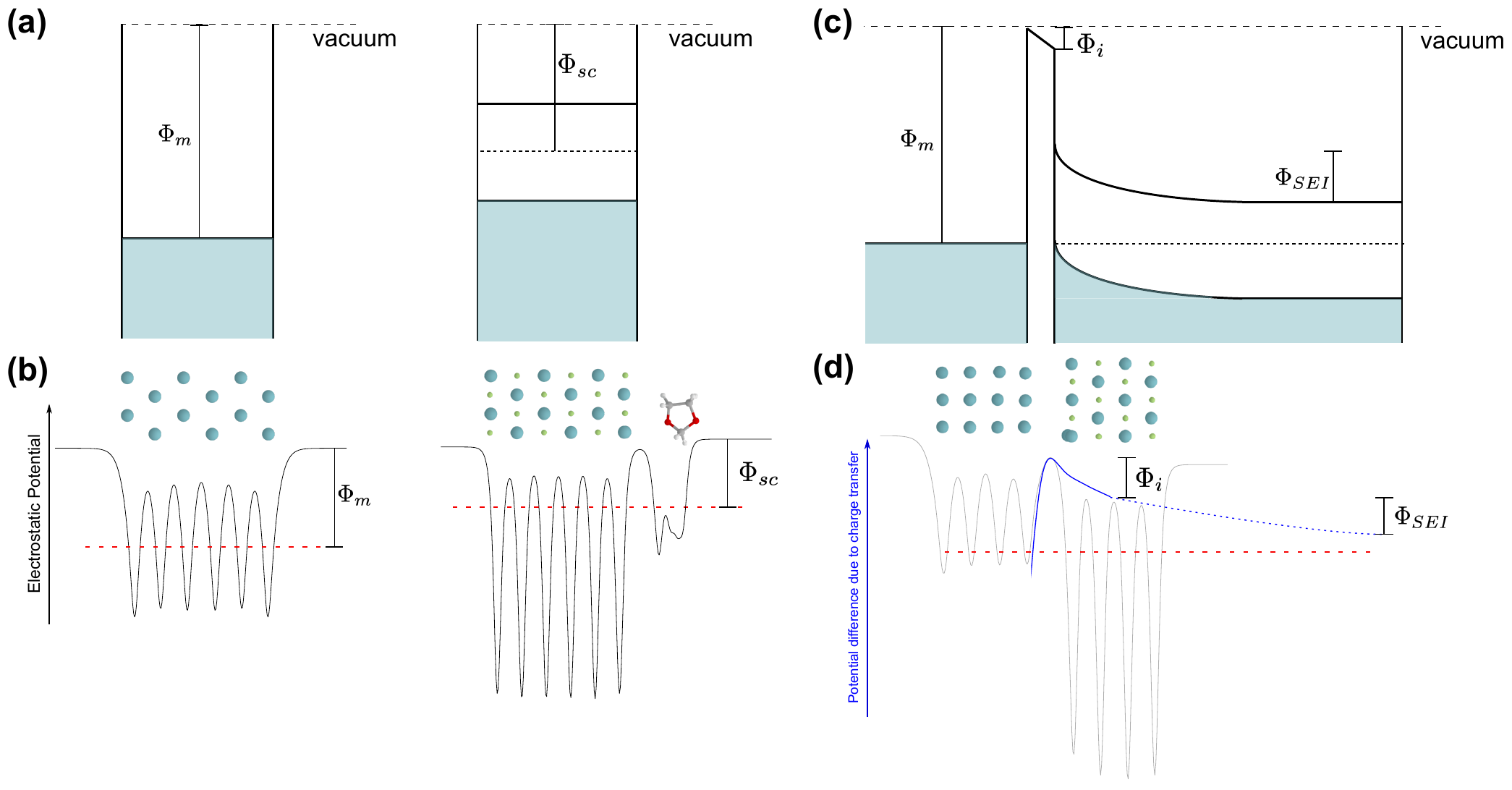}
    \caption{Schematic diagram of the potential drops associated with the formation of a Schottky barrier. (a) Before the metal and insulator are brought into contact, they each have separate work functions. To equalize the work functions within the materials when they are placed in contact (b), a potential gradient is formed consisting of a portion due to charge states directly at the interface as well as a portion due to band bending within the insulator. The top half of the diagrams shows the traditional schematics for Schottky barrier formation, while the bottom half shows the equivalent DFT systems as well as their planar averaged potentials. It should be noted that the potentials illustrated here, while coming directly from relevant DFT calculations, are not to scale with each other. Furthermore, the difference of work function and degree of band bending is exaggerated in this diagram to illustrate the relevant concepts.}
    \label{fig:schottky-barrier}
\end{figure}

We believe the most useful analogy for understanding the potential landscape in a battery interface is the Schottky barrier in electronics.
When a metal and semiconductor/insulator meet (in this case, the anode and the SEI, respectively), we expect that an electronic barrier will form \cite{tung2001recent}.
As illustrated in Fig.~\ref{fig:schottky-barrier}a, the work function, \textit{i.e.} the potential needed to move an electron from the Fermi level of the material to vacuum, of the isolated metal $\Phi_{m}$ and semiconductor $\Phi_{sc}$ will likely differ. 
When these two components interact, charge will be transferred until the Fermi levels of the components are in equilibrium, creating one Fermi level for the entire system. 

Metals have orders of magnitude higher charge carrier concentration than semiconductors \cite{ashcroft2022solid}, so charge can be moved from the metal without substantially altering the metal's Fermi level.
Because of the limited charge carrier concentration within a semiconductor, however, the transferred charge can be spread up to several hundreds or thousands of nm, which leads to substantial `band bending' within the semiconductor. 
The exact nature of this band bending can be determined by applying the Poisson equation to Boltzmann statistics for a semiconductor.
Exact solutions have been worked out for a number of relevant cases  ~\cite{myamlin1963electrochemistry,gerischer1968semiconductor,rajeshwar2007fundamentals}.
The magnitude of the potential barrier created by this band bending is termed the ``Schottky barrier'' within the electronics community. 
In this work, we will refer to it as the SEI barrier, $\Phi_{SEI}$, as this is the semiconducting/insulating layer in contact with the metal in the battery context.

In a perfect interface, the SEI barrier will exactly match the difference between the metal and semiconductor work functions:
\begin{equation}
    \Phi_{SEI} = \Phi_m - \Phi_{sc}.
\end{equation}
Real world interfaces will not be perfect, however, and charge traps and other defects will tend to accumulate at the metal-semiconductor interface. 
These interfacial states lead to an additional potential barrier $\Phi_i$, which will typically reduce the total SEI barrier. 
We can then write the SEI barrier as 
\begin{equation}
    \Phi_{SEI} = \Phi_m - \Phi_{sc} - \Phi_{i}.
    \label{eq:sei_barrier_int}
\end{equation}
When voltage is applied to the system, this can be taken as altering the effective work function of the metal 
\begin{equation}
    \Phi_m' = \Phi_m + e_0 V_{app}
\end{equation} 
where $V_{app}$ is the applied voltage.
We can then finally determine the voltage dependent SEI barrier as 
\begin{equation}
    \Phi_{SEI}(V_{app}) = \Phi_m + e_0 V_{app} - \Phi_{sc} - \Phi_{i}(V_{app}).
\end{equation}
Note that the interfacial potential is also voltage dependent.

Determining the exact magnitude of $\Phi_i$ versus $\Phi_{SEI}$ for a given system and voltage requires a quantum mechanical simulation of the specific interface placed in equilibrium with the Poisson-Boltzmann potential-charge distribution within the SEI.
We outline how this is achieved in the following section. 

\subsection{The Quantum Continuum Approximation (QCA) for determining voltage on a Li ion SEI interface}

To calculate the first principles response of Li ion transport in Li ion batteries, we use QCA for metal-insulator-electrolyte interfaces, the full details of which have been previously described by Campbell \cite{campbell2023voltage}.
This QCA methodology was first developed for semiconducting-electrolyte interfaces \cite{campbell2017quantum}, and has demonstrated excellent agreement with experimental characterization for Schottky barriers in metal-semiconductor interfaces \cite{subramanian2020photophysics}.

In brief, explicit DFT calculations of both the metal-insulator and insulator-electrolyte interfaces along with regions of interest, such as atoms surrounding Li ion transport, are coupled to continuum representations of the Poisson-Boltzmann charge-voltage distribution within a bulk insulator. 
For a perfectly insulating/semiconducting system, the charge distribution can be well approximated by the following Poisson--Boltzmann relationship:\cite{schmickler2010interfacial}
\begin{equation}
    \frac{d^2\phi}{dz^2} = \frac{n_{\text{d}}}{\epsilon_o \epsilon_{\text{sc}}}  \left[ 1 - \exp \left( \frac{-\phi}{k_{\text{B}}T}\right)\right].
    \label{eq:pb-ms-conditions}
\end{equation}
To incorporate the impacts of voltage, a total charge $Q$ for the entire electrode is selected, and then multiple distributions of charge between the anode, SEI, and surface are calculated.
To balance these included charges, we use Helmholtz planes of countercharge.
Each of these differing charges introduces different electric fields to each section distinct DFT interface calculation.
We can then quantify the potential distribution of the bulk SEI region using the Poisson-Boltzmann conditions outlined in Eq.~\ref{eq:pb-ms-conditions}. 
The potential within the SEI region can then be approximated as:
\begin{equation}
    \bar{\phi}^{\text{SEI}}(z) = \bar{\phi}(z_{\text{edge}}^{\text{SEI}|\text{sol.}}) + \frac{\epsilon_o \epsilon_{\text{SEI}}}{2n_{\text{d}}}\left[\left(\frac{d\bar{\phi}^{\text{SEI}}}{dz}(z)\right)^2 - \left(\frac{d\bar{\phi}}{dz}(z_{\text{edge}}^{\text{SEI}|\text{sol.}})\right)^2\right] - k_{\text{B}}T ,
    \label{eq:ox-pot}
\end{equation}
where $\bar{\phi}$ is the planar averaged potential in the system due to the inclusion of voltage/charge, $k_{\text{B}}$ is the Boltzmann constant, and $T$ is the temperature of the system. 
This relationship applies throughout the bulk oxide region, which lasts from $z_{\text{edge}}^{\text{ox.}|\text{sol.}}$ to $z_{\text{edge}}^{\text{m}|\text{ox.}}$.
With Eq.~\ref{eq:ox-pot}, we can then predict the voltage offset $\Phi_{\text{SEI}}$ within the SEI.
We then find the equilibrium distribution of charge by finding the distribution of charge between the anode, SEI, and surface that causes the Fermi level to stay constant throughout the entire system:
\begin{equation}
    \varepsilon_{\text{F}}^{\text{SEI.}|\text{sol.}} = \varepsilon_{\text{F}}^{\text{m}|\text{SEI}} + \Phi_{\text{SEI}},
\end{equation}
where $\varepsilon_{\text{F}}^{\text{SEI}|\text{sol.}}$ is the DFT Fermi level of the SEI-solution interface, and $\varepsilon_{\text{F}}^{\text{m}|\text{SEI}}$ is the DFT Fermi level of the metal anode-SEI interface. 
Since the countercharges within the model take the form of Helmholtz planes, this gives a Helmholtz model for the electrical double layer within the electrolyte. 
This provides a useful first approximation of the electrical double layer response, but in the future, this could be replaced with a Guoy-Chapman or similar improved model. 

From this we can then use the Trassati relation \cite{trasatti1986absolute} to extract an equilibrium electronic anode potential $\phi_e$ for the system as 
\begin{equation}
    \phi_e = -E_{F}/|e| - 1.37  \text{ V (Li/Li}^+),
\end{equation}
where $|e|$ is the electronic charge, assuming that the potential far away from the electrode has been set to zero.
This offset of 1.37 is based on the standard electrode potential for a Li/Li$^+$ reaction of a Li anode in an aqueous solution as compared to SHE\cite{haynes2016crc}.
In principle, the exact chemical makeup of the electrolyte would change this value, but we use this value as a useful first approximation, particularly for interpreting results relative to each other.
$\phi_e$ represents the electrochemical potential of one of of the halves of our battery system, in the case of our examples, the anode. 
This anode potential can be related to the open circuit voltage of the battery system $V_{oc}$ by measuring the difference between the cathode potential $\phi_e^{\text{cathode}}$ and the anode potential $\phi_e^{\text{anode}}$, 
\begin{equation}
    V_{oc} = \phi_e^{\text{cathode}} - \phi_e^{\text{anode}}.
    \label{eq:batt_voltage}
\end{equation}

We note that this is an electronic definition of voltage, which is typically used throughout the computational electrochemistry community \cite{cheng2012alignment}.
This differs from the widespread definition of ionic voltage $\phi_i$ within the battery DFT community \cite{aydinol1997ab}, which is based on the energy of inserting or removing Li atoms from the simulation 
\begin{equation}
    \phi_i = \left[(E_{n} - E_{n-1}) - \mu_{Li}\right]/|e|,
\end{equation}
where $E_n$ is the DFT energy of the system with $n$ Li atoms and $\mu_{Li}$ is the chemical potential of bulk Li. 
These definitions of voltage do not necessarily lead to the same results (and, in fact, often do not). 
Leung has defined $\phi_i$ as the ``equilibrium'' voltage, which depends on Li ions fully equilibrating within the system to environmental conditions, and $\phi_e$ as the ``instantaneous'' voltage, which depends on the orders of magnitudes faster movement of electrons \cite{leung2020dft}.
For a system where $\phi_i \neq \phi_e$, we can interpret the system as being at an overpotential. 
Throughout this work we will focus on the electronic potential that an interface is under.
Future work will focus on further exploration of the interactions between these different potential definitions. 

\subsection{QCA for voltage dependent diffusion barriers}
\label{subsec:qca_rxn_barriers}

To calculate how the electronic potential impacts the diffusion barrier of a given Li ionic transport reaction, we first calculate the transition state using standard DFT Nudged Elastic Band (NEB) methodology. 
This methodology generated several images along the transition pathway from the initial to the final state.
This transition pathway is then optimized and the image with the highest energy, termed the climbing image, is found and the reaction barrier calculated as the difference between the climbing state energy and the initial state energy. 
With the climbing image identified, we calculate how the energy of the both the climbing image and the starting state of the reaction change as a function of voltage using QCA. 
The electronic potential specific diffusion barrier $\Delta E_{b}$ at a specific electronic potential $\phi_e$ can thus be calculated as 
\begin{equation}
    \Delta E_{b}(\phi_e) = E^{CI}(\phi_e) - E^{IS}(\phi_e),
\end{equation}
where $E^{CI}(\phi_e)$ and $E^{IS}(\phi_e)$ are the DFT calculated energy of the climbing image and initial state, respectively, using QCA to control for the electronic potential.
This approach mimics a similar approach recently taken by Vijay \textit{et al.} to calculate potential dependent reaction barriers for molecular catalysis on metallic surfaces \cite{vijay2022force}.
Due to computational restrictions, we assume that the climbing image remains the same across different potentials, which is largely consistent with the results from Vijay \textit{et al.}\cite{vijay2022force}

\subsection{QCA and Computational Details}
The exact thickness and defect concentration of a given SEI are highly dependent on the anode, the electrolyte environment, and cycling conditions \cite{wang2018effect}.
Throughout this work, we will assume a temperature of 300 K, a SEI thickness of 200 \AA, and a charge carrier concentration of $10^9$ cm$^{-3}$ as representative values \cite{shi2013defect,veith2015direct}. 
The exact voltages and electric fields reported in this work and their impact on Li ion transport throughout the SEI are necessarily dependent on these values. 
However, since we are working with relatively low charge carrier concentrations (at least compared to typical semiconducting systems), we do not see significant band bending as might be expected in a more semiconducting system (at least on the length scale of a few hundred \AA ). 
Instead, the potential continues in an essentially straight line throughout the SEI, following the slope established at the surface until meeting the secondary interface with the metal. 
We thus see little change in the reported barriers from increasing or decreasing the charge carrier concentration by a few orders of magnitude. 
These results are only applicable for the charge carrier concentration and SEI lengths we examine here. 
Due to computational restrictions, we cannot examine all possibilities; this does mean, however, that these results may not be representative all operational conditions of a battery. 
This issue is particularly exacerbated by the fact that the charge carrier concentration may not be uniform throughout a realistic SEI.
Future work will focus on developing more localized sections of charge density and ionic pathways to reflect this mix.
This would result in changes to Eq.~\ref{eq:ox-pot} and the need for more explicit calculations at regions of interest. 
It may be reasonable to treat current results as representing an average charge carrier concentration across a broad area, however.
In this paper, we solely examine cases where the anode is a pure Li metal, but our methodology is broadly applicable to any anode and we expect that the general conclusions about the dependence of Li ion transport barriers on the electronic potential and SEI chemistry of the battery will remain largely the same.

All electronic structure calculations are done using the {\sc quantum espresso} package~\cite{giannozzi2009quantum}.
We use norm-conserving pseudopotentials from the PseudoDojo repository~\cite{van2018pseudodojo} and the Perdew-Burke-Ernzerhof exchange-correlation functional~\cite{perdew1996generalized}.
We use kinetic energy cutoffs of 50 Ry and 400 Ry for the plane wave basis sets used to describe the Kohn-Sham orbitals and charge density, respectively.
We use a 2$\times$2$\times$1 Monkhorst-Pack grid~\cite{monkhorst1976special} to sample the Brillioun zone in our calculations.
Interface structrues were generated using the pymatgen interface generator \cite{ong2013python}.
Grain boundary structures are generated using the AIMSGB algorithm and toolset \cite{cheng2018aimsgb}. 
Equilibrium structures are found by allowing the forces to relax below 0.05 eV/\AA.

We use the {\sc Environ} package \cite{andreussi2012revised} to calculate the parabolic corrections to the surface dipole as well as including the planes of Helmholtz charges needed for QCA.
For simplicity, all calculations are done in vacuum with only a first monolayer of dioxolane (DOL) molecules included as part of the electrolyte, which has been shown to improve band alignment of DFT calculations \cite{hormann2019absolute}.
In the supporting information, we demonstrate the dependence of the calculated work function on the number of DOL layers included, finding reasonable variability with the number of included DOL layers, although our monolayer results are within 0.2 eV of a three layer structure, indicating this is at least useful as a first approximation of the electrolyte system. 
Exact atomic coordinates and structural parameters used for all calculations are included in the supporting information. 

\section{Results}\label{sec2}
\subsection{SEI barriers}
Using the QCA methodology established in Sec.~\ref{sec:methods}, we simulate the SEI potential barrier as a function of voltage in LiF, Li$_2$O, Li$_2$CO$_3$, as well as a LiF-LiF and LiF-Li$_2$O grain boundary SEI systems. 
For each of these systems, we report the interfacial barrier as well, which, as outlined above, is key to determining the total potential drop across the SEI as a function of voltage, and thus for impacting Li ion transport through the SEI.

\subsubsection{LiF}
\label{subsec:LiF}

\begin{figure}
    \centering
    \includegraphics[width=\columnwidth]{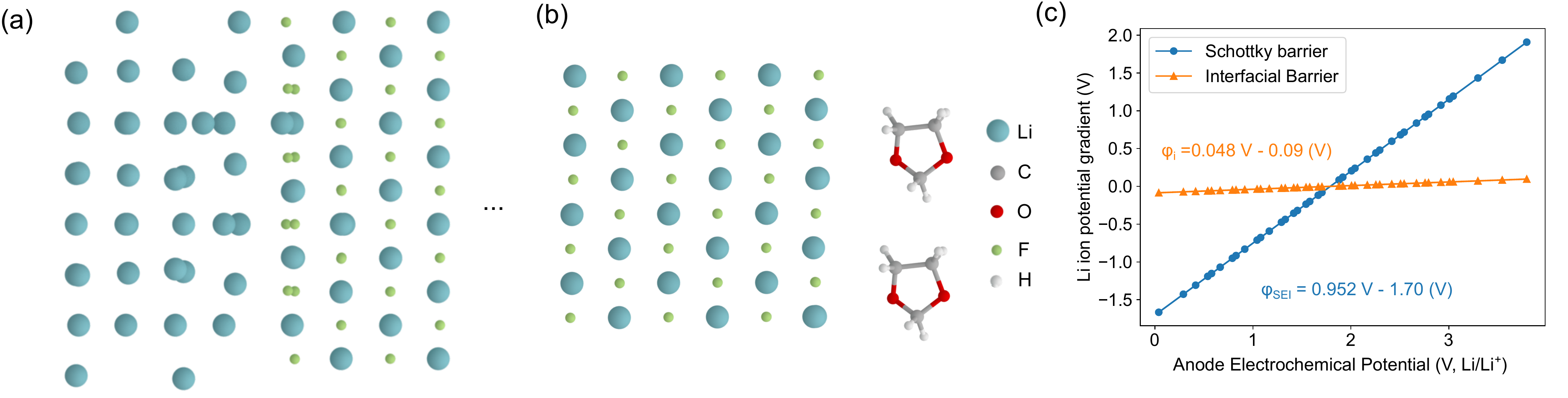}
    \caption{(a) The Li/LiF interface modeled, along with (b) the LiF surface with a subsaturation coverage of DOL electrolyte molecules. (c) The resulting potential barriers within the SEI pushing or pulling Li ions at the surface toward the anode. Negative barriers indicate that the Li ion is thermodynamically encouraged to move ``downhill'' toward the anode.}
    \label{fig:lif-barrier}
\end{figure}
We first examine the SEI barrier of LiF, where the interfacial anode-SEI structure is shown in Fig.~\ref{fig:lif-barrier}a and the SEI-electrolyte structure shown in Fig.~\ref{fig:lif-barrier}b.
We calculate both the SEI barrier that a Li ion will face moving from the electrolyte through the SEI to the anode surface $\Phi_{SEI}$ as well as the interfacial barrier a Li ion will need to overcome in moving through the anode-SEI interface $\Phi_{i}$ in Fig.~\ref{fig:lif-barrier}c. 
Both these values are dependent on the overall voltage of the system and are zero at the 'flatband' voltage of the system, which in this case corresponds to 1.88 V (Li/Li$^+$).
Away from the flatband voltage, the barrier is linearly responsive to the change in voltage, as will be true for all of the systems examined in this work. 

A linear relationship between the SEI barrier and the voltage is not guaranteed for any given system, and is dependent on the extent of the band bending within the SEI. 
In this case, we are assuming a charge carrier concentration of 10$^9$ cm$^{-3}$ and a length of 20 nm, which, while relatively typical in batteries, are both relatively low relative to the typical semiconductor interface. 
This means that the electrostatic potential ends up exhibiting essentially no band bending within the SEI, and the potential profile in space is, in fact, linear. 
This constant slope of the potential within the SEI results in a linear relationship between the voltage and the potential for these systems. 
A higher charge concentration (e.g. 10$^{14}$ cm$^{-3}$) and/or a longer length SEI would result in a more pronounced band bending, which would partially break this linear relationship between the voltage and the SEI barrier magnitude. 

Within this framework, we can then extract the slope of the SEI barrier $\Phi_{SEI}$ and the metal-interfacial barrier $\Phi_{i}$ as a function of the voltage using linear regression, with the exact equations shown in Fig.~\ref{fig:lif-barrier}c.
We can see that the majority of the potential drop for any given voltage ($\approx$95 \%) happens within the SEI with only a small amount of metal-SEI interfacial barriers ($\approx$ 5 \%).
This indicates that the Li-LiF interface is relatively clean in this example, producing only a small amount of interfacial charge trapping. 

From the perspective of a Li ion moving from the electrolyte to the anode at a given electronic voltage, it will have to overcome (or be helped by) the SEI potential barrier over the 20 nm of the bulk SEI width. 
We take the convention that a positive SEI barrier implies that the Li ion will have to overcome the potential barrier, and a negative barrier implies that the Li ion will be able to move ``downhill'' with respect to the electrostatic potential, \textit{i.e.}, be encouraged to move toward the anode. 
Given the positive slope of the SEI barrier as a function of voltage, our results imply that it will become easier and easier for the Li ion to migrate to the anode as the anode electrochemical potential decreases.
This matches experimental findings of Li plating becoming stable as the anode electrochemical potential decreases \cite{lin2021lithium}.
It further supports naive intuitions of battery charging becoming easier at higher open circuit battery voltages $V_{oc}$, as can be seen by lowering the anode electrochemical potential in Eq.~\ref{eq:batt_voltage}. 
Notably, the point at which the SEI barrier encourages Li migration versus discouraging it is highly interface specific (in the case of the Li-LiF structure studied here, it occurs at 1.88 V (Li/Li$^+$)). 
We next go on to see how the relationship between these barriers changes with a differing SEI composition.

\subsubsection{Li$_2$O}

\begin{figure}
    \centering
    \includegraphics[width=\columnwidth]{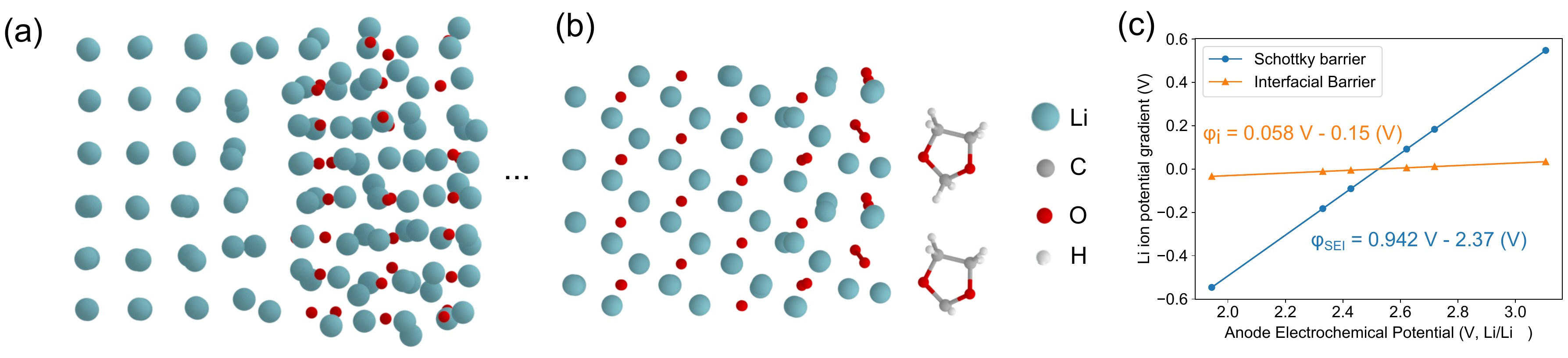}
    \caption{(a) The Li/Li$_2$O interface modeled, connected to (b) the Li$_2$O surface with a subsaturation coverage of DOL electrolyte molecules. (c) The resulting potential barriers within the SEI pushing or pulling Li ions at the surface toward the anode. }
    \label{fig:li2o-barrier}
\end{figure}
We predict the SEI and anode-SEI interfacial barrier for Li transport through a Li$_2$O SEI as a function of voltage in Fig.~\ref{fig:li2o-barrier}.
In many ways, the barriers here resemble what we have already seen for the LiF SEI. 
The SEI barrier dominates the response of the system to electronic voltage, absorbing $\approx$94\% of the voltage drop on the system.
This does represent a slight increase in the barrier distributed at charge traps at the Li/Li$_2$O interface, rising to $\approx$6\%.
This rise can be interpreted as the density of charge traps slightly increasing in comparison to LiF.

An important difference, however, is the anode potential at which the SEI barrier switches from encourage Li transport from the electrolyte to the anode to encouraging this transfer.
For the Li$_2$O SEI, this threshold potential is 2.53 V (Li/Li$^+$)), nearly 0.6 V higher than what we saw for LiF.
This implies that a Li battery with a Li$_2$O SEI will encourage Li transport toward the anode at higher anode potentials (and thus, by Eq.~\ref{eq:batt_voltage}, lower open circuit voltages) than a battery with a LiF SEI.
Depending on the application of interest, this could potentially be used to increase the energy density of a charged battery, or conversely to lower the voltage requirements for full charging.

\subsubsection{Li$_2$CO$_3$}
\begin{figure}
    \centering
    \includegraphics[width=\columnwidth]{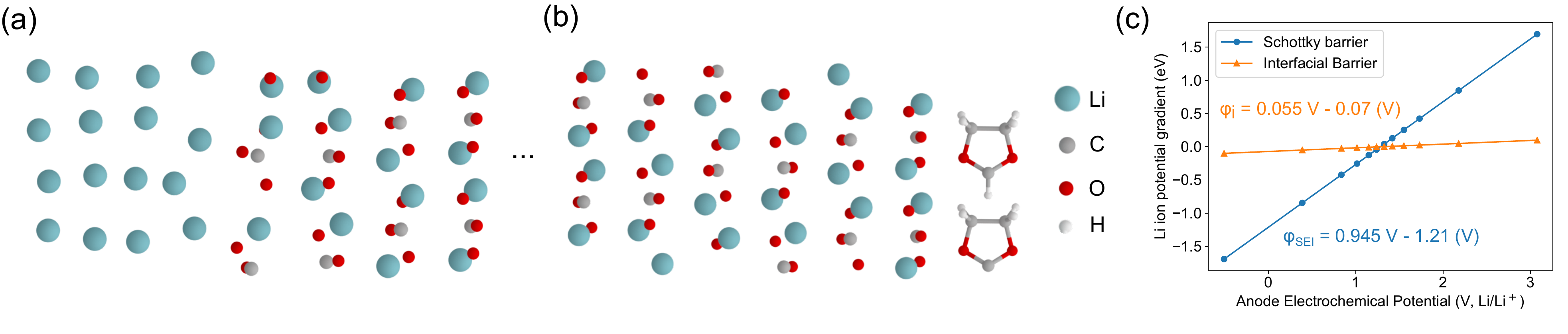}
    \caption{ (a) The Li/Li$_2$CO$_3$ interface modeled, connected to (b) the Li$_2$CO$_3$ surface with a subsaturation coverage of DOL electrolyte molecules. (c) The resulting potential barriers within the SEI pushing or pulling Li ions at the SEI surface toward the anode. }
    \label{fig:li2co3-barrier}
\end{figure}
We predict the SEI and anode-SEI interfacial barrier for Li transport through a Li$_2$CO$_3$ SEI as a function of voltage in Fig.~\ref{fig:li2co3-barrier}.
For the Li$_2$CO$_3$ SEI, the threshold voltage of switching from a negative barrier encouraging Li ion transport toward the anode to a positive barrier discouraging transport is 1.28 V (Li/Li$^+$)), nearly 0.6 V lower than what we saw for LiF.
The nature of the surface states remain largely the same though, with the bulk of the barrier ($\approx$94\%) taking place across the bulk SEI.
Only ($\approx$6\%) of the barrier is captured at surface states. 

For these fundamentally clean interfaces, it seems that the exact surface chemistry does not lead to large changes in the amount of charge trapped at the anode-SEI interface. 
It does, however, play a substantial effect on determining the threshold voltage at which the SEI barrier switches from encouraging Li transport toward the anode versus away from the anode. 
Depending on the desired application, it may be desirable to adjust the SEI chemistry to encourage charging at higher or lower voltages.
\subsubsection{Grain boundaries}
\begin{figure}
    \centering
    \includegraphics[width=\columnwidth]{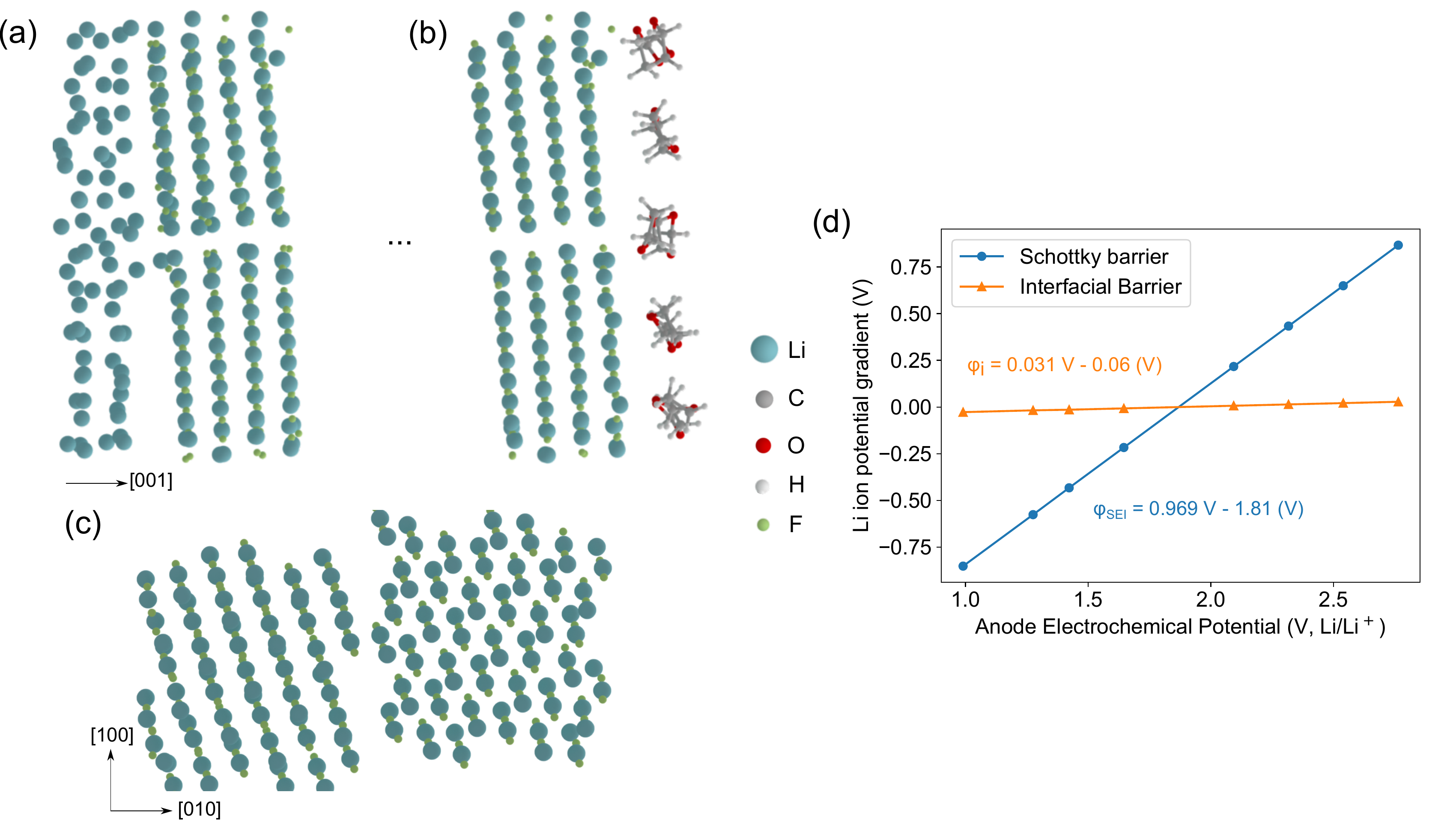}
    \caption{(a) The Li/LiF-LiF interface modeled, connected to (b) the LiF-LiF surface with a subsaturation coverage of DOL electrolyte molecules. (c) A front view of the LiF-LiF grain boundary, demonstrating the channels through which Li ions may flow. (d)The resulting potential barriers within the SEI pushing or pulling Li ions at the surface toward the anode. }
    \label{fig:lif-lif-barrier}
\end{figure}

We next examine a $\Sigma$5 LiF-LiF grain boundary on a Li metal anode. 
Grain boundaries are expected to exhibit significant impacts on the overall transport of Li ions through the SEI, with larger grain boundaries potentially serving as channels for ion transport \cite{ramasubramanian2019lithium}.
It is thus important to analyze how the SEI barrier may change at these grain boundaries as they may represent the rate determining barrier that  Li ions face in charge/discharge operations.
The atomic configuration of the LiF-LiF grain boundary we simulate is shown in Fig.~\ref{fig:lif-lif-barrier}. 

We apply QCA and predict the SEI barrier as a function of electronic voltage, as shown if Fig.~\ref{fig:lif-lif-barrier}d.
We find that the proportion of the total potential barrier taken up by interfacial states ($\approx$3\%) versus the proportion that is distributed throughout the SEI ($\approx$97\%) is almost identical to the split in the interfacial barrier for a pure crystalline LiF SEI. 
Similarly, the voltage threshold at which the potential barrier switches from helping Li ion progress toward the anode to hindering progress is almost identical at 1.87 V (Li/Li$^+$)). 
This demonstrates that the surface state properties which determine the potential barrier within the SEI are largely independent of any grain boundaries that may (and in fact likely do) exist between two chemistries that are the same.

\begin{figure}
    \centering
    \includegraphics[width=\columnwidth]{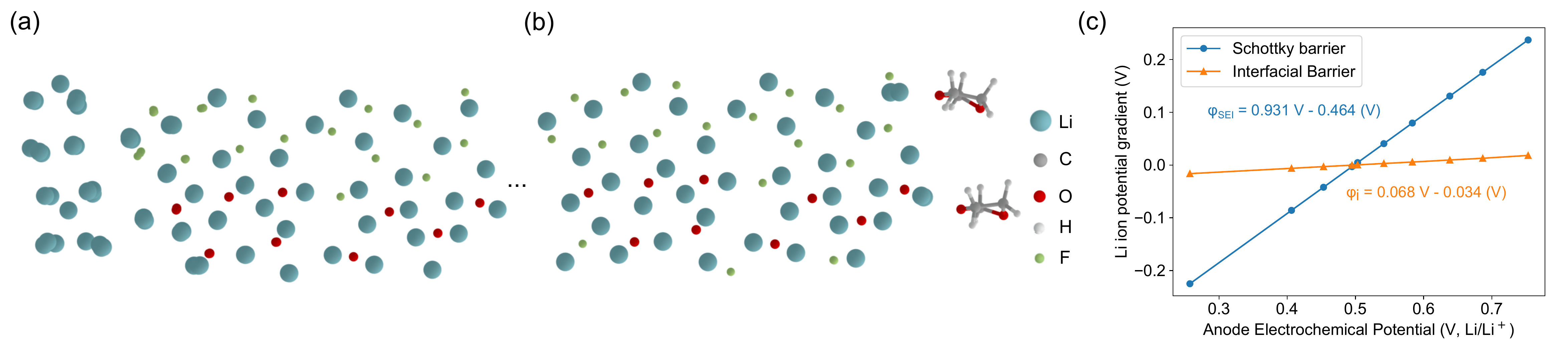}
    \caption{(a) The Li/LiF-Li$_2$O interface modeled, connected to (b) the LiF-Li$_2$O surface with a subsaturation coverage of DOL electrolyte molecules. (c) The resulting potential barriers within the SEI pushing or pulling Li ions at the surface toward/away from the anode. }
    \label{fig:lif-li2o-barrier}
\end{figure}

Finally, we examine an SEI composed of differing chemistries: a LiF-Li$_2$O grain boundary, with the atomic structure shown in Fig.~\ref{fig:lif-li2o-barrier}a-b. 
We find that the proportion of the total barrier taken up by the interfacial charge trap states ($\approx$7\% as shown in Fig.~\ref{fig:lif-li2o-barrier}c) matches that of Li$_2$O relatively well.
The SEI barrier similarly composes $\approx$93\% of the total barrier. 
This indicates that in a grain boundary with multiple chemical compositions, the level of interfacial charge trapping will be largely determined by the composition with the highest interfacial trapping. 
The voltage threshold of crossover between encouraging Li ion transport from the electrolyte toward the anode, however, is 0.50 V (Li/Li$^+$)). 
The SEI barrier therefore favors Li transport toward the anode (charging) near these grain boundaries only at significantly lower anode potentials than either LiF (1.93 V Li/Li$^+$) or Li$_2$O (2.53 V Li/Li$^+$) single composition SEIs.
At lower potentials, this translates to larger SEI potential drops favoring Li transport toward the anode.
For instance, at a voltage of 0.2 V (Li/Li$^+$), our calculations lead us to predict a SEI potential drop of -1.51 V in a LiF SEI, -2.15 V in Li$_2$O, and  only -0.28 V in the modeled LiF-Li$_2$O grain boundary.
This indicates that the specific LiF-Li$_2$O grain boundary modeled makes migration of the anode significantly harder, with larger open circuit voltages required to achieve similar levels of SEI assistance to Li ion transport.

\subsection{Li diffusion barriers}

While these SEI barriers can clearly amount to significant amounts over the length scale of the entire SEI, it is important to examine how the potential drop will practically impact diffusion barriers for Li transport across an individual lattice site at any given location. 
\subsubsection{Li transport in LiF}
\begin{figure}
    \centering
    \includegraphics[width=\columnwidth]{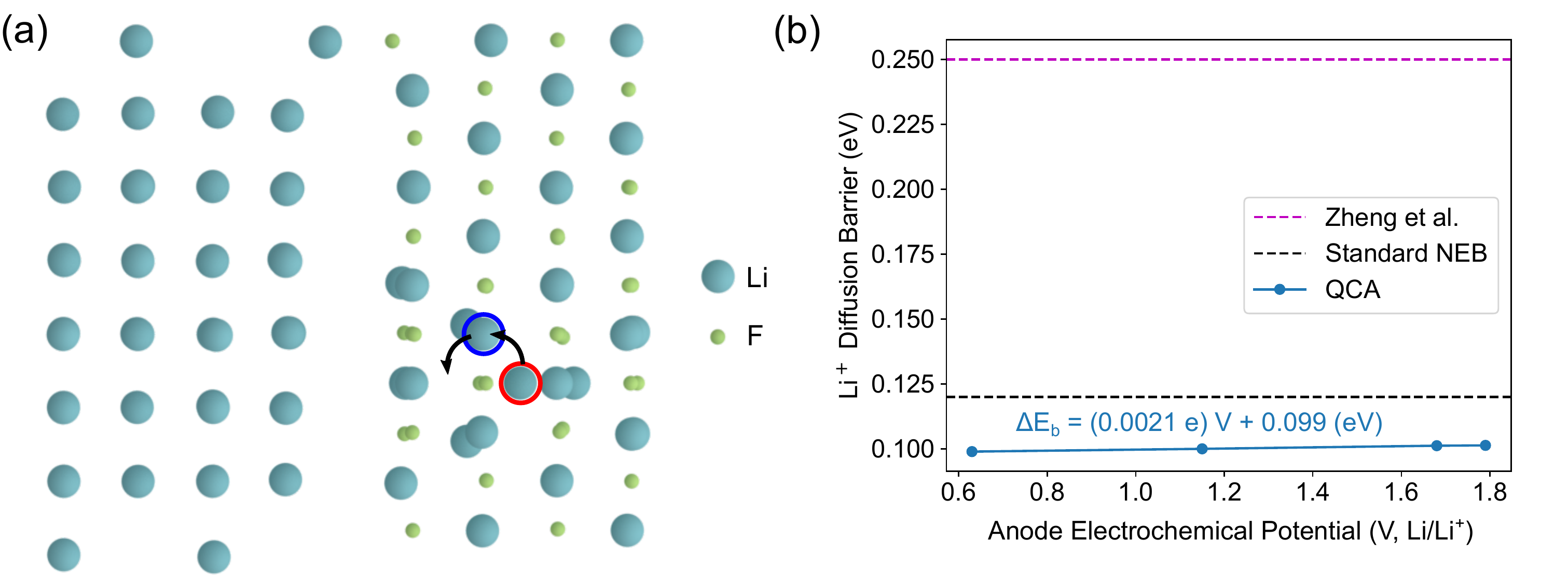}
    \caption{(a) The reaction path for a Li interstitial knock off mechanism moving through a LiF SEI. (b) The Li diffusion barrier as a function of voltage. Given the small distance the Li ion is traveling for this path, the actual potential difference that is conferred from the SEI barrier is limited. The literature diffusion barrier comes from Zheng \textit{et al.}\cite{zheng2021lithium}. }
    \label{fig:lif-rxn-barrier}
\end{figure}

We simulate the knock-off mechanism for interstitial movement in a LiF SEI, as depicted in Fig.~\ref{fig:lif-rxn-barrier}a.
We then change the potential of both the starting and transition state with QCA to find the potential dependent barrier, as shown in Fig.~\ref{fig:lif-rxn-barrier}b.
This approach differs from previous work that examines Li interstitial barriers that only looks at neutral bulk conditions and thus ignores the contributions of voltage such as Zheng \textit{et al.}, who find a barrier of 0.25 eV for this reaction \cite{zheng2021lithium}. 
Using standard DFT NEB calculations, we predict a 0.12 eV barrier.
This is already significantly different and may be explained by our inclusion of a Li-metal anode in addition to the LiF.
Further, the QCA derived potential of both the transition state and the starting state of the reaction are not kept constant during typical NEB calculations. 
This means we need to adjust the potential of both images to match before subtracting the energies to derive the QCA barriers, as described in Sec.~\ref{subsec:qca_rxn_barriers}. 

We find that the reaction barrier changes minimally with changes in the electronic voltage and corresponding SEI barrier. 
With regression, we predict a reaction barrier magnitude $\Delta E_b$ as a function of voltage as 
\begin{equation}
    \Delta E_b = (0.0021 e) \phi_e + 0.099 eV,
    \label{eq:rxn-barrier}
\end{equation}
where $\phi_e$ is the anode electrochemical potential of the system with respect to the Li/Li$^+$ standard electrode. 
This voltage dependence is fairly unimpressive! 
However, it largely tracks with the local distribution of the SEI barrier as this reaction barrier is for a Li ion traveling a much smaller distance than the entire SEI and thus will face only a fraction of the total SEI barrier.
As a concrete example, at a voltage of 0.2 V (Li/Li$^+$), we expect to see a SEI barrier of $\Phi_{SEI} =$-1.51 eV across the entire 200 \AA\ LiF SEI favoring Li ion transport to the anode.
The interstitial Li ion in the reaction shown in Fig.~\ref{fig:lif-rxn-barrier}a is traveling $\approx$0.5 \AA\ in the direction perpendicular to the interface, and we can thus expect that the portion of the SEI barrier it encounters over this period would proportionally be -0.0038 eV.
When we predict the diffusion barrier for this system using Eq.~\ref{eq:rxn-barrier}, we predict that the diffusion barrier will be lowered by -0.0035 eV from its ``neutral'' value at the flatband potential, nearly an exact match.

We can thus say that the impact of altering the voltage of the electrode SEI will be minimal for any given Li diffusion barrier over a few \AA.
There will, however, be a small alteration of the diffusion barrier corresponding to the magnitude of the SEI barrier at a given voltage divided by the fraction of the SEI length that the reaction takes place (assuming the charge carrier concentration is similar to what we model here, resulting in linear potential distributions). 
Furthermore, while the voltage based reduction of the barrier for any given Li movement across one site is minimal, this reaction needs to be repeated hundreds of times to cross the entire SEI.
This means that the total amount of energy needed for a Li ion to move across the entire SEI can be significantly reduced (or increased) by changing the electronic voltage. 

There is currently no experimental evidence to clearly distinguish between the reaction barriers shown due to the difficulty of individual atomic measurements. 
Future work using careful \textit{in-situ} analysis with tools such as Kelvin Probe Force Microscopy and Quartz Crystal Microbalance measurements could be useful for helping narrow the range of theoretical uncertainty, as well as demonstrating how these barriers change as a function of voltage. 
We believe that our QCA methodology represents a higher accuracy method for determining these reaction barriers in realistic battery contexts as we control for the electronic voltage of the calculation. 

\subsubsection{Li transport in LiF-LiF grain boundary}

\begin{figure}
    \centering
    \includegraphics[width=\columnwidth]{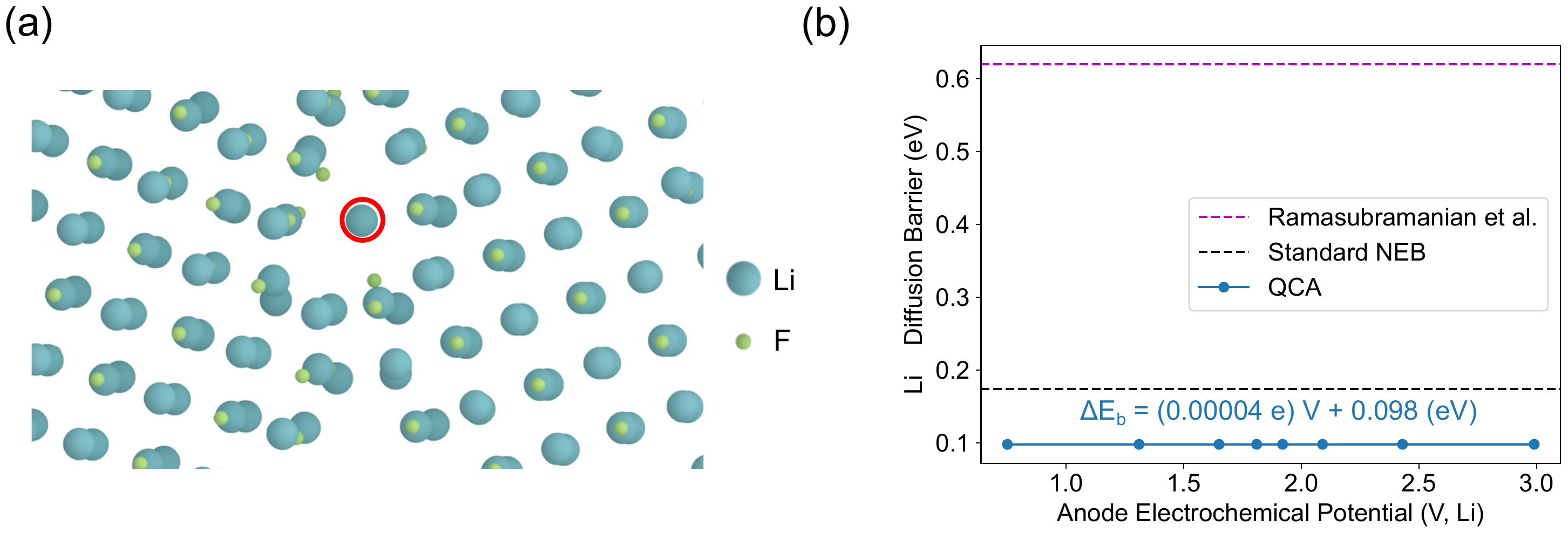}
    \caption{(a) The reaction path for a Li moving toward the anode through the grain boundary. We highlight the moving Li in red. The Li ion is moving into the page toward the anode. (b) The Li diffusion barrier as a function of voltage. The literature diffusion barrier comes from Ramasubramanian \textit{et al.}\cite{ramasubramanian2019lithium}}
    \label{fig:lif-lif-rxn-barrier}
\end{figure}
Finally, we examine the reaction barrier for a Li ion moving through the channel of the $\Sigma$5 LiF-LiF grain boundary explored earlier. 
This reaction barrier additionally includes the interface with the Li anode, which can provide a key differentiating factor. 
This approach contrasts with previous work of Ramasubramanian \textit{et al.} \cite{ramasubramanian2019lithium}, which finds the barrier for Li diffusion through this grain boundary to be 0.68 eV when examining the SEI in isolation.
Using standard NEB calculations where we include the Li anode interface, we predict a diffusion barrier of 0.17 eV, which is already significantly reduced. 
Once again, we then extend this analysis with QCA of the initial and climbing image to determine the diffusion barrier as a function of voltage. 
Using a linear regression, we predict the diffusion barrier magnitude as
\begin{equation}
    \Delta E_b = (0.00004 e) \phi_e + 0.098 eV.
    \label{eq:rxn-barrier-lif-lif}
\end{equation}
Once again, the dependence on voltage appears fairly minimal, although the inclusion of electronic voltage control lowers the overall barrier by $\approx$0.07 eV.

Our work highlights that previous work that has not been including both interfaces with anodes and electronic voltage may significantly overestimate reaction barriers for Li ion transport. 
This overestimation seems largely driven by the lack of inclusion on an anode-SEI interface within the reaction barrier calculations. 
Altering the voltage on the material, however, can have impacts on the total energy needed for a Li to diffuse across the full SEI, and represents a useful tool for tuning a given devices performance. 
This demonstrates that some reaction barriers which had previously been theoretically predicted to be too high for realistic operation may actually be feasible when controlling for interface and voltage effects within the calculation. 

\section{Discussion}\label{sec:discussion}

Our simulations are necessarily imperfect representations of real battery devices.
While QCA extends the length scale examined perpendicular to the surface, it does not extend the lateral area explored, which is treated using typical DFT periodic boundary conditions. 
This means that simulations of individual SEIs such as LiF and Li$_2$O inherently assume that only these perfectly crystalline chemistries exist in the SEI.
In reality, the surface is known to likely be a mixture of different SEI chemistries interacting with each other. 
While our simulations of grain boundaries begin to address this concern, they are necessarily still periodic and are thus assuming a repeating pattern of grain boundaries across the surface, often with quite high surface densities. 
Our work will serve as a stepping stone, however, for future multiscale research aiming to address the larger lateral surface.

For the sake of demonstrating the concept in this work, the interfacial structures studied were determined from relaxation of the underlying interface using DFT, but did not account for atomic movement that may happen during thermalization. 
To fully account for these factors, \textit{ab initio} Molecular Dynamics (AIMD) simulations, ideally at elevated temperatures, should be used to generate baseline configurations of the interface for the follow--on QCA calculations of barrier height. 
Nevertheless, these interfaces provide useful lower bounds on the amount of charge trapping we can expect for a given SEI, i.e. this represents the minimum charge trapping in the case of an essentially a perfect interface.

Another aspect that limits the potential real world accuracy of our calculations is the representation of the electrolyte. 
For ease of demonstrating QCA within these systems, we only simulated rather shallow monolayers of DOL electrolyte molecules along with the inorganice SEI, but organic SEI components based on reactions with the electrolyte are often observed as well \cite{schechter1999x}. 
This will be addressed with future work including a realistic mix of electrolyte molecules and a full Guoy-Chapman-Stern analytic inclusion of the charge distribution. 
QCA represents a tractable method for including as much information about electrolytes as feasible into these calculations before resorting to relatively expensive AIMD.

Finally, our work is predicated on the SEI behaving largely as an insulator. 
The input charge carrier concentration is assumed to be uniform throughout the system, but may also be viewed as an average concentration for these system.
This does not, however, capture when there is significant local variation in this concentration, which determines whether a system is insulating or (semi)conducting. 
Future work will focus on integrating regions of varied conductivity into our calculations at the expense of additional DFT SEI calculations. 

Despite these limitations, the methodology used here represents a substantial improvement in the ability of DFT battery calculations to include electronic voltage and to calculate relevant barriers for Li transport. 
This has allowed us to provide novel information on the how the SEI barrier and individual reaction barriers change as a function of voltage and SEI composition, quantities that cannot be easily experimentally accessed. 

\section{Conclusion}
Our QCA analysis reveals that the barrier for transporting Li through the SEI of a battery is dependent on the overall applied electronic voltage, and to  large extent on the location of the Li ion in the SEI. 
This implies that controlling the chemical composition of the SEI and applied voltage may be a useful tool for modulating the rate of charging for a given battery systems. 
In general, this trend matches intuition as higher open circuit voltages lead to lower transport barriers, favoring easier charging.
The threshold voltage at which Li ion movement is encouraged toward the anode versus away from the anode, however, is highly dependent on the specific SEI chemistry.
We see that, at least in some cases, grain boundaries lower this potential threshold more than would be seen for the individual components. 
We find that controlling for interface and electronic voltage effects can significantly lower the predicted Li diffusion barriers when compared to literature.
This implies that some previous reactions which had been dismissed as requiring too much energy may be plausible when correctly accounting for these effects. 
Furthermore, it is likely that as models of the anode-SEI interface become more realistic, this effect will become more pronounced, as evidenced by the interface's significant impact on our QCA calculations. 

Future work will focus on utilizing similar types of calculations as part of a cluster expansion theory that could be used to extend the lateral length scale examined and find equilibrium SEI structures as a function of potential. 
Once these equilibrium structures are found, they could be the input for a e.g. Kinetic Monte Carlo code that uses the barriers in this work to predict actual rates of Li ion transport as a function of potential. 
This work thus represents the first step in a multiscale methodology for using first principles to predict battery device behavior with SEI chemistries and voltage dependent transport rates that resemble real world systems. 

This analysis shows that QCA can be a useful tool for identifying optimal charging chemistry and voltage conditions for a given device. 
While ionic potentials (i.e. equilibrium potentials of adding or removing a Li to a system) are frequently calculated using DFT, we are unaware of other techniques to clearly identify charging potentials in real world, out of equilibrium contexts.

\begin{suppinfo}
	Supporting information is available providing: (1) input files with the atomic structures of all systems calculated throughout this work, 
	and (2) calculation of the dependence of the calculated work function on the number of explicit solvent layers included.
\end{suppinfo}

\begin{acknowledgement}
We would like to thank Kevin Leung and Peter Schultz for useful discussions as well as reviewing a draft version of this manuscript.

This work was supported by the Laboratory Directed Research and Development program at Sandia National Laboratories under project 229741.
Sandia National Laboratories is a multi-mission laboratory managed and operated by National Technology \& Engineering Solutions of Sandia, LLC (NTESS), a wholly owned subsidiary of Honeywell International Inc., for the U.S. Department of Energy’s National Nuclear Security Administration (DOE/NNSA) under contract DE-NA0003525. This written work is authored by an employee of NTESS. The employee, not NTESS, owns the right, title and interest in and to the written work and is responsible for its contents. Any subjective views or opinions that might be expressed in the written work do not necessarily represent the views of the U.S. Government. The publisher acknowledges that the U.S. Government retains a non-exclusive, paid-up, irrevocable, world-wide license to publish or reproduce the published form of this written work or allow others to do so, for U.S. Government purposes. The DOE will provide public access to results of federally sponsored research in accordance with the DOE Public Access Plan
\end{acknowledgement}

\bibliography{refs_new}

\end{document}


\setcounter{section}{0}
\setcounter{page}{1}
\renewcommand{\thepage}{S\arabic{page}} 
\renewcommand{\thesection}{S\arabic{section}}  
\renewcommand{\thetable}{S\arabic{table}}  
\renewcommand{\thefigure}{S\arabic{figure}}

\section*{Dependence of Calculated Work Function on DOL layers}
\label{sec:Work-fct}
\begin{figure}
    \centering
    \includegraphics[width=0.5\columnwidth]{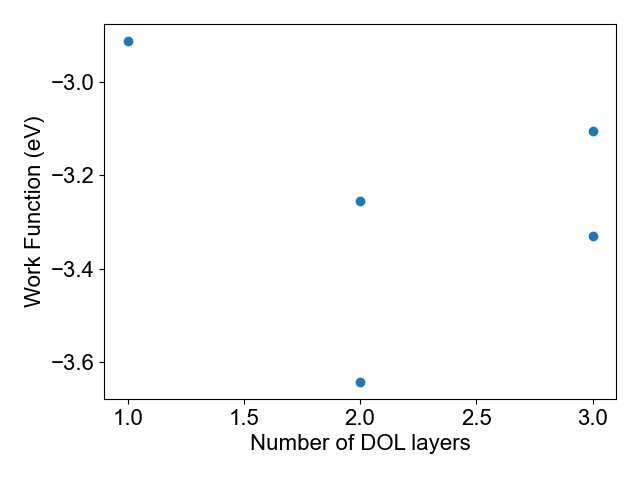}
    \caption{Dependence of the work function on the number of DOL layers simulated for a LiF SEI. }
    \label{fig:work-fct}
\end{figure}
In the main text, all of our calculations are performed with a monolayer of DOL molecules to simulate the interaction of the surface with the electrolyte.
This single monolayer has been shown to improve band alignment of DFT calculations \cite{hormann2019absolute}.
It is important to quantify, however, how well this approximation captures the work function of a system with more complete simulations of multiple DOL layers. 
In Fig.~\ref{fig:work-fct}, we examine the dependence of the work function on the number of DOL layers that we model. 
The calculation used within the main manuscript is kept as the data point for 1 layer of DOL molecules.
We then simulate two different configurations of DOL layers for both two total layers and three total layers. 
For each of these systems, we see a wide spread in the calculated work function, demonstrating the importance of these molecules for work function determination.
The ideal approach would be to run an AIMD simulation of a significant amount of molecules and then take an averaged work function from several snapshots within that run. 
Our results show that the monolayer of DOL represents an adequate first approximation of the system, with a work function within 0.2 eV - 0.4 eV of the high end three layer calculations we carry out.
This emphasizes that the usefulness of our results for relative prediction between similar systems.

\bibliography{refs_new}